\documentclass[11pt,twoside]{article}

\usepackage{asp2006}
\usepackage{epsf}
\usepackage{psfig}
\usepackage{lscape}
\usepackage{graphicx}

\begin{document}

\title{Sunspot umbra atmosphere from full Stokes inversion}   								
\author{R.~Wenzel\altaffilmark{1}, S.V.~Berdyugina\altaffilmark{2}, D.M.~Fluri\altaffilmark{1}, J.~Arnaud\altaffilmark{3}, and A.~Sainz-Dalda\altaffilmark{4}}   	
\altaffiltext{1}{Institute for Astronomy, ETH Zurich, 8093 Zurich, Switzerland}
\altaffiltext{2}{Kiepenheuer-Institut f\"ur Sonnenphysik, 79104 Freiburg, Germany}
\altaffiltext{3}{LUAN, Universit\'e de Nice, Nice 06108, France}
\altaffiltext{4}{Stanford-Lockheed Institute for Space Research, USA}

\begin{abstract} 
Sunspots are prominent manifestations of the solar cycle and provide key constraints for understanding its operation. Also, knowing internal structure of sunspots allows us to gain insights on the energy transport in strong magnetic fields and, thus, on the processes inside the convection zone, where solar magnetic fields are generated and amplified before emerging at the surface on various scales, even during solar minima.
In this paper, we present results of a spectropolarimetric analysis of a sunspot observed during the declining phase of the solar cycle 23. By inversion of full Stokes spectra observed in several spectral regions in the optical at the THEMIS facility we infer the height dependence of physical quantities such as the temperature and the magnetic field strength for different sunspot regions. The simultaneous use of atomic (Fe~{\sc i} 5250.2 and 5250.6 \AA) and highly temperature sensitive molecular (TiO 7055 \AA\ and MgH 5200 \AA) lines allow us to improve a model of the sunspot umbra.
\end{abstract}


\section{Introduction}   

Inversion of spectropolarimetric data becomes a standard tool for recovering the vertical structure of solar magnetic features, including sunspots \citep[e.$\,$g.,][]{mathew2003,beck2008}. However, in most cases the inversion is limited to atomic lines formed in a narrow range of heights in the solar atmosphere. Simple diatomic molecules like MgH and TiO are sensitive to both the magnetic field and the temperature and provide information on different atmospheric layers of the solar photosphere than atomic lines \citep{berdyugina2002, berdyugina2003}. They can be observed in the sunspot umbra at typical temperatures below 4500\,K. Combining spectropolarimetry of atomic and molecular lines thus provides a very sensitive tool for detecting and probing magnetic structures in the sunspot umbra, which so far was done for infrared Fe~{\sc i} and OH lines \citep{mathew2003} and for optical Fe~{\sc i}, MgH, and TiO lines \citep{afram2006}. In this paper, we further employ the advantage of this approach. 

%
%
\section{Observations}

\begin{figure}[!ht]
    \begin{center}
	  \includegraphics[width=6cm, angle=-90]{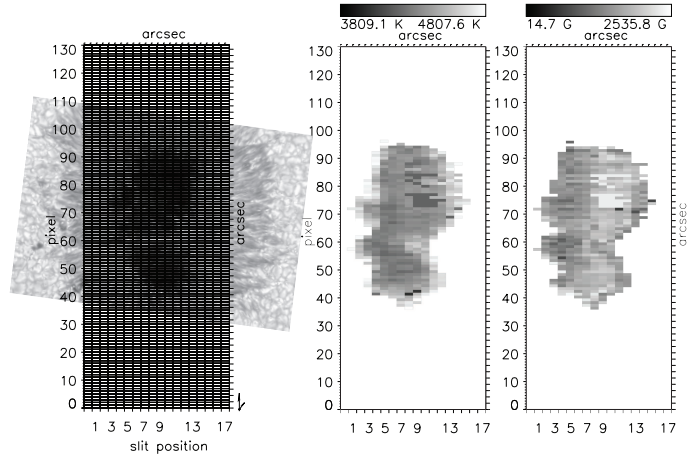}
    \end{center}
  \caption{{\it Left}: Image of the sunspot NOAA 10667 obtained in the TiO 7055 \AA\ band at the SST. The resolution and position of the THEMIS data are indicated by the grid. The top and right hand side scales give the angular size of the sunspot which is about 15$\times$25 arcsec$^2$. For each pixel the full Stokes vector was recorded. {\it Right}: The results of the inversion. These maps show the temperature and magnetic field stratifications at log $\tau_{5000}$ = $-$1 for the cool second magnetic component.}
  \label{fig:spot}
  \end{figure}

The sunspot NOAA 10667 was observed at the THEMIS facility (Tenerife) during several consecutive days in September 2004 \citep{afram2006, arnaud2006}. Full Stokes profiles were recorded simultaneously in five spectral regions. From the available data three spectral windows were chosen for the present analysis: Fe~{\sc i} lines at 5250.2 \AA\ and 5250.6 \AA, TiO at 7055 \AA, and MgH at 5200 \AA. As indicated by contribution functions shown in Fig.~\ref{fig:model}, the chosen atomic and molecular lines carry information on different layers in the sunspot atmosphere, where they constrain our model. In addition, their different Land\'e factors allow us to determine both the strength of the magnetic field and the magnetic filling factor \citep{stenflo1973}. 
This sunspot was also imaged during a coordinated program at the SST (La Palma) with a narrow-band filter centered at the TiO 7055 \AA\ band, as designed by \citet{berger_berdyugina2003} (see Fig.~\ref{fig:spot}).

%
%


\section{Inversions}

\begin{figure}[!ht]
   \begin{center}
	  \includegraphics[width=12cm]{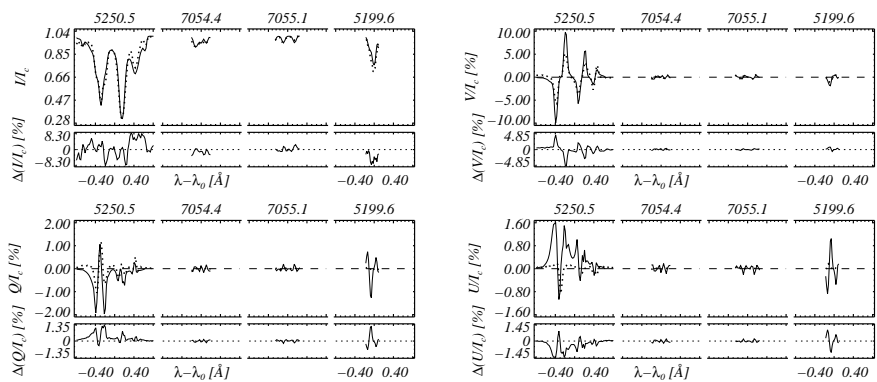}
   \end{center}
  \caption{The best fit to a typical set of Stokes profiles for one umbral pixel located at (7, 53) according to the grid shown in Fig.~\ref{fig:spot}. The dotted line is observational data, the solid line is the synthetic spectrum found by inversion.}  
  \label{fig:fits} 
\end{figure}

Stokes parameters {\it I/I}$_{\rm c}$, {\it Q/I$_{\rm c}$}, {\it U/I$_{\rm c}$} and {\it V/I$_{\rm c}$} in the selected spectral regions were inverted for each pixel independently employing the code SPINOR \citep{frutiger2000, berdyugina2003}. This code assumes LTE conditions and solves the Unno-Rachkovsky radiative transfer equations using response functions.
Starting from an initial guess the Stokes profiles are fitted iteratively to the observational data by varying parameters of the model atmosphere. In Fig.~\ref{fig:fits} an example is shown when temperature and magnetic field strength are iterated as height dependent free parameters at five reference points log $\tau_{5000}$ = $-$4, $-$3, $-$2, $-$1, and 0. 

After extensive tests with different atmospheric models, a three component atmosphere was found to give the best fit results. The first component is a standard quiet sun atmosphere accounting for straylight, whereas the second and third component represent magnetic parts of the sunspot atmosphere. 
All spectral windows and thus all atomic and molecular lines were inverted simultaneously.

\nopagebreak
\section{Results}

\begin{figure}[!ht]
  \begin{center}
	\includegraphics[width=7cm]{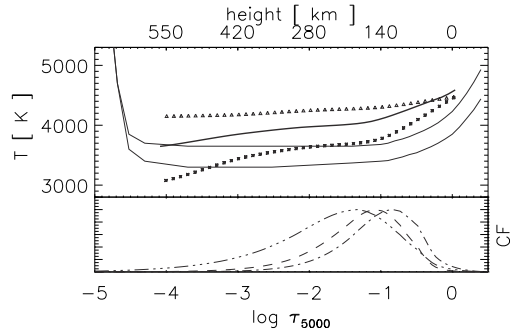}
  \end{center}
\caption{A comparison of different umbra models. Average umbral temperature profiles for all components (bold line) and the two magnetic components separately (triangles and boxes) are presented together with models of \citet{maltby1986} (thin solid lines, models E and L from top to bottom). An indication of the depths probed by our method is given below where the normalized line depression contribution functions (CF) are plotted for Fe~{\sc i} (dash triple dotted), TiO (dashed), and MgH (dot dashed) lines.}
  \label{fig:model}
\end{figure}

Even though the inversion was applied to the whole set of data, here we discuss results for the sunspot umbra only. As indicated above, the resulting atmosphere leading to best fits consists of three components. The overall magnetic filling factor is about 0.92 splitted almost equally between the two magnetic components. These components have different temperature gradients near the bottom of the atmosphere (see Fig.~\ref{fig:model}) and most probably indicate the presence of
unresolved umbral structures such as umbral dots imbedded into a diffuse dark umbra. 

We compare our results with models of the dark core umbra by \citet{maltby1986}. Those are also shown in Fig.~\ref{fig:model} and correspond to early (E) and late (L) phases of the solar cycle or may be interpreted as having the properties of dark and bright umbra, respectively. Clearly, our umbral models are warmer than those of Maltby. This is in agreement with a systematic increase of the umbral temperature during at least the last decade as reported by \citet{penn_livingston2006}.

\section{Conclusions}

The aim of this study is to improve a sunspot model atmosphere. Simultaneous full Stokes inversions were carried out employing atomic and molecular lines such as Fe~{\sc i}, TiO, and MgH.  This allowed us to increase the range of temperature sensitivity and improve the accuracy in depths ranging from log $\tau =-$4 to log $\tau \leq$0. As compared to existing umbral temperature profiles \citep{maltby1986}, the sunspot we analyzed can be best modeled with a warmer atmosphere consisting of three components accounting for unresolved umbral structures. 

A detailed discussion of this sunspot model atmosphere is in preparation \citep{wenzel2010}.

\acknowledgements 
This work is supported by the SNF grant PE002-104552. We thank V. Zakharov for providing the image of the sunspot. SVB acknowledges the EURYI Award from the ESF (www.est.org/euryi).




\end{document}